\documentclass{moriond}

\usepackage{wrapfig}
\usepackage{float}
\def\Journal#1#2#3#4{{#1} {\bf #2}, #3 (#4)}

\def\PLB{{\em Phys. Lett.}  B}

\def\PRD{{\em Phys. Rev.} D}

\def\ra{\rightarrow}

\def\be{\begin{equation}}
\def\ee{\end{equation}}
\def\bea{\begin{eqnarray}}
\def\eea{\end{eqnarray}}

\def\ttbar{$t\bar{t}$}
\def\ifb{fb$^{-1}$}
\def\pt{$p_{\textrm{\tiny{T}}}$}
\def\Haa{$H\ra aa$ }

\newcommand{\Photo}{\includegraphics[height=35mm]{picture}}

\begin{document}
\vspace*{4cm}
\title{HIGGS RARE AND EXOTIC DECAYS}

\author{LJILJANA MORVAJ \footnote{On behalf of the ATLAS and CMS
    Collaborations} \footnote{Copyright 2021 CERN for the benefit of
    the ATLAS and CMS Collaborations. Reproduction of this article or
    parts of it is allowed as specified in the CC-BY-4.0 license.}}

\address{CERN, Esplanade des Particules 1,\\
1217 Meyrin, Switzerland}


\maketitle\abstracts{
Several recent searches for exotic and rare decays of the Standard
Model Higgs boson with the ATLAS and CMS detectors are presented. The
searches are performed on $\sqrt{s}$=13 TeV proton-proton
collisions data collected at the LHC between 2015 and 2018. The topics covered include
 searches for the Higgs boson decays into two pseudoscalars,
 $H\rightarrow aa$, in three different final states,
 $2b2\mu$, $4b$ and $2\mu2\tau$, search for lepton-flavour violating
 Higgs decays, $H\rightarrow  \mu\tau / e\tau$, and search for a
 rare $H\rightarrow \ell\ell \gamma$ decay.
ATLAS presents evidence for the $H\rightarrow
\ell\ell \gamma$ rare decay, amounting to an observed
significance of $3.2\sigma$.
}

\section{Introduction}

Rare and exotic Higgs boson decays searches provide important
tests of the Standard Model (SM) and could lead to
a discovery of new physics. 
Current fits to the SM Higgs couplings constrain the branching fractions of
the Higgs boson to undetected, $B(H\ra$ undetected), and
invisible, $B(H\ra$ invisible), final states to be less than 19\%
and 9\%, respectively~\cite{ATL_comb}. This leaves a lot of space for new physics in
the Higgs  boson decays. 
This talk presents several recent results delivered by the ATLAS
\cite{ATL} and CMS~\cite{CMS} collaborations using 13 TeV
proton-proton collision data collected at the LHC.
Three types of searches are covered: Higgs decays into
beyond-the-SM (BSM) states ($H\rightarrow aa \rightarrow 2b2\mu / 4b /
2\mu2\tau$), lepton flavor violating Higgs decays ($H\rightarrow \mu\tau/e\tau$) and rare Higgs decays
($H\rightarrow \ell\ell \gamma$).

\section{Higgs decays to pseudoscalars}
Light bosons appear in many well-motivated extensions of the Standard
Model. For example, they could be mediators between the SM and some hidden sector that does not interact
through the weak, strong or electromagnetic forces~\cite{Curtin}.
Searches for light pseudoscalars in Higgs boson decays are often
interpreted in Two-Higgs-Doublets plus a Singlet (2HDM+S) set of
models~\cite{Curtin}. The pseudoscalar component of the singlet field ($a$) acquires Yukawa-like couplings to SM particles through mixing with
Higgs bosons. This implies that the largest branching fractions of the
$a$-boson will generally be to $b-$quarks and $\tau-$leptons. 
For $a$-masses ($m_a$) that are significantly lower than the SM Higgs boson mass, the
$a$-boson is boosted and its decay products are collimated. 
Merging of  $b$-jets and hadronically decaying $\tau$-leptons  into one reconstructed object starts below roughly
$m_a$=25 GeV.
ATLAS and CMS have developed dedicated reconstruction
and identification techniques to address such boosted final states.

\subsection{$H\rightarrow aa \rightarrow 2b2\mu$}
\label{subsec:Hbbmm}
The final state with two $b$-jets and two muons~\cite{H2b2m_ATL}
provides a good
balance between the typically large decay fraction of $a\ra bb$ and a
clean signature of a narrow dimuon resonance from the $a\ra \mu\mu$
side. This feature of the decays is further exploited in a
kinematic-likelihood (KL) fit. In the fit, the di-$b$-jet mass ($m_{bb}$) is
constrained, within its resolution, to the dimuon mass ($m_{\mu\mu}$) measured with
approximately ten times better resolution. The output of the KL fit is
the maximum likelihood score, which can be used to select 
events satisfying the $m_{bb} \sim m_{\mu\mu}$ hypothesis with high
efficiency for the signal, while rejecting a large portion of SM backgrounds. A Boosted
Decision Tree (BDT) discriminant is trained to discriminate the
signal against the dominant backgrounds consisting of top-quark pair (\ttbar) and
Drell-Yan (DY) production. A number of kinematic variables
characterising event topology are used in the BDT training. For
example, small (large) angular separation between the two $b$-jets
(between the dimuon and the di-$b$-jets systems) characterises the signal
topology at lower $m_a$, while the opposite is true for higher
$m_a$. The BDT training is done separately at each signal mass in order to
maximally exploit mass-dependent signal characteristics and maximise
the sensitivity across the considered $m_a$ range. The dimuon
invariant mass spectrum is scanned in 2 GeV or 3 GeV wide bins in
search for an excess above the background model. The background
expectations and the observed data in all the bins are shown in Figure~\ref{fig:bbmm1}. The largest deviation
is observed at 52 GeV and corresponds to a local (global) significance
of 3.3$\sigma$ (1.7$\sigma$). The upper limits on $B(H\rightarrow aa \rightarrow
2b2\mu)$ are shown in Figure~\ref{fig:bbmm2}, left, in red (observed) and black
(expected). The limits are compared to the result from an earlier ATLAS
publication based on 36 \ifb of data~\cite{H2b2mOld_ATL} (blue lines) and show factors 2--5
improvement over the full mass range.

\begin{figure}[H]
\centering
\includegraphics[width=0.8\linewidth]{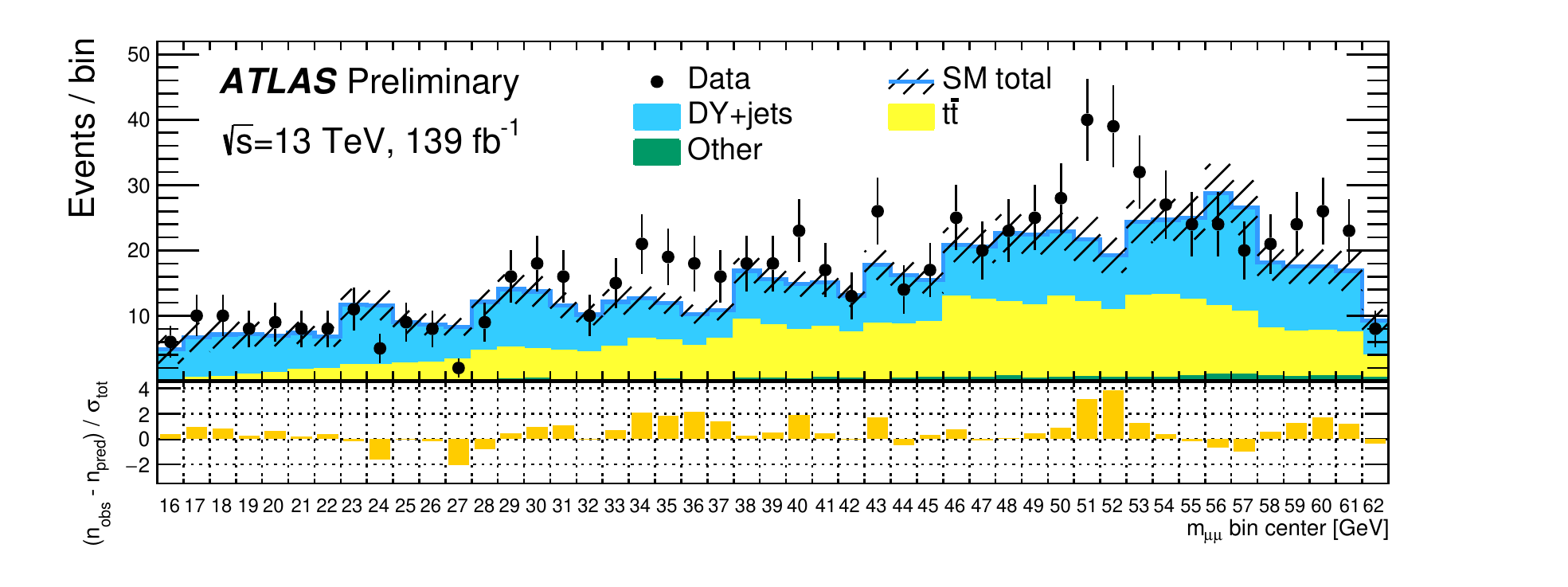}
\caption{Observed data and estimated backgrounds in all $m_{\mu\mu}$
  bins that are tested for the presence of signal in $H\rightarrow aa
  \rightarrow 2b2\mu$ analysis \protect\cite{H2b2m_ATL}. The bins are
  2 GeV (3 GeV) wide in $m_{\mu\mu}$ for $m_a\leq45$~GeV ($m_a>45$~GeV). Events in neighbouring bins partially overlap. 
The bottom panel shows the pull in each bin, defined as $(n_\mathrm{obs}-n_\mathrm{pred})/\sigma_\mathrm{tot}$, where $n_\mathrm{obs}$ is the number of events in the data, $n_\mathrm{pred}$ is the number of the fitted background events and $\sigma_\mathrm{tot}$ is the total (systematic and statistical, added in quadrature) uncertainty on the fitted background yield.}
\label{fig:bbmm1}
\end{figure}

\begin{figure}
\includegraphics[width=0.49\linewidth]{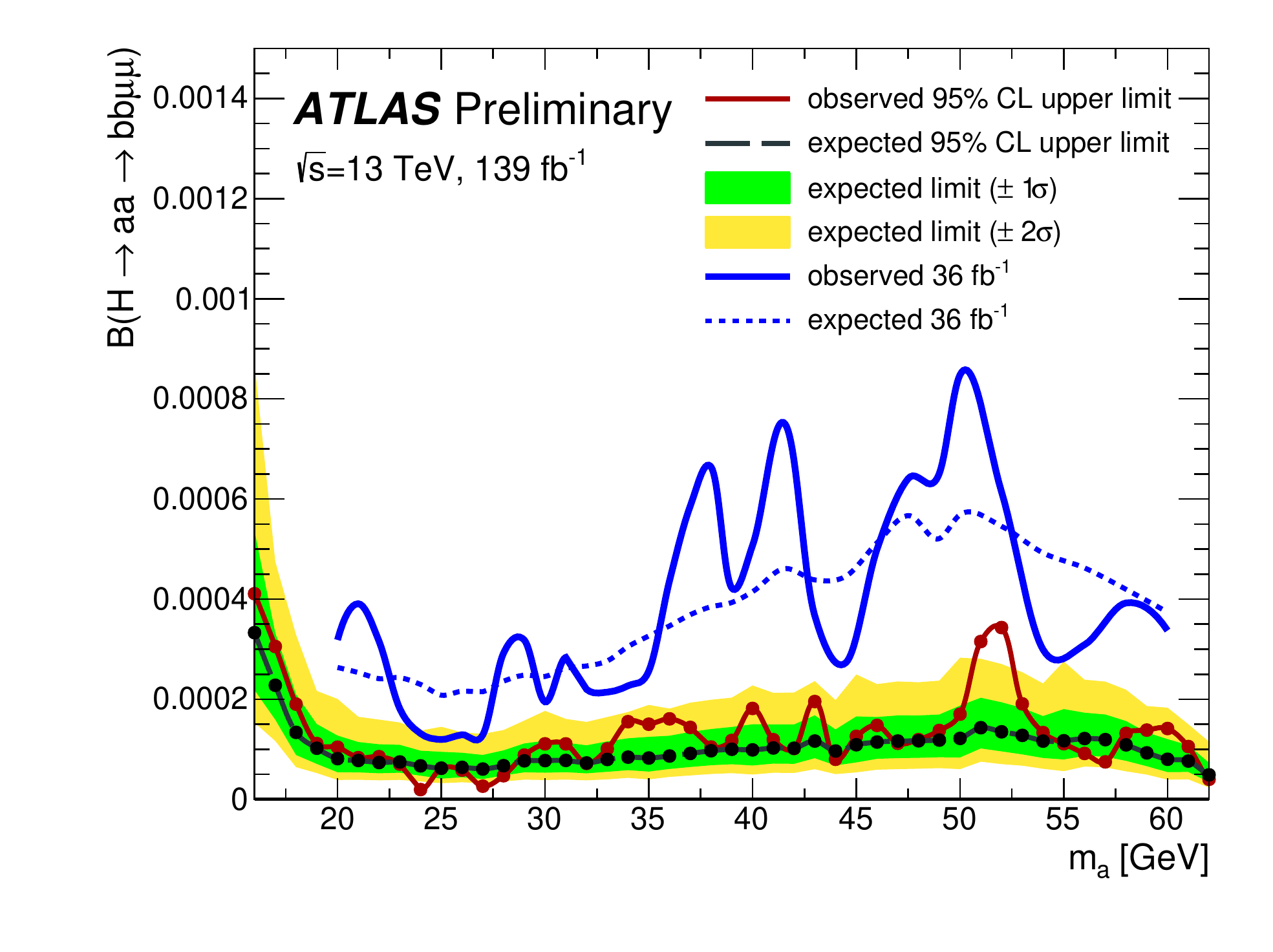}
\includegraphics[width=0.49\linewidth]{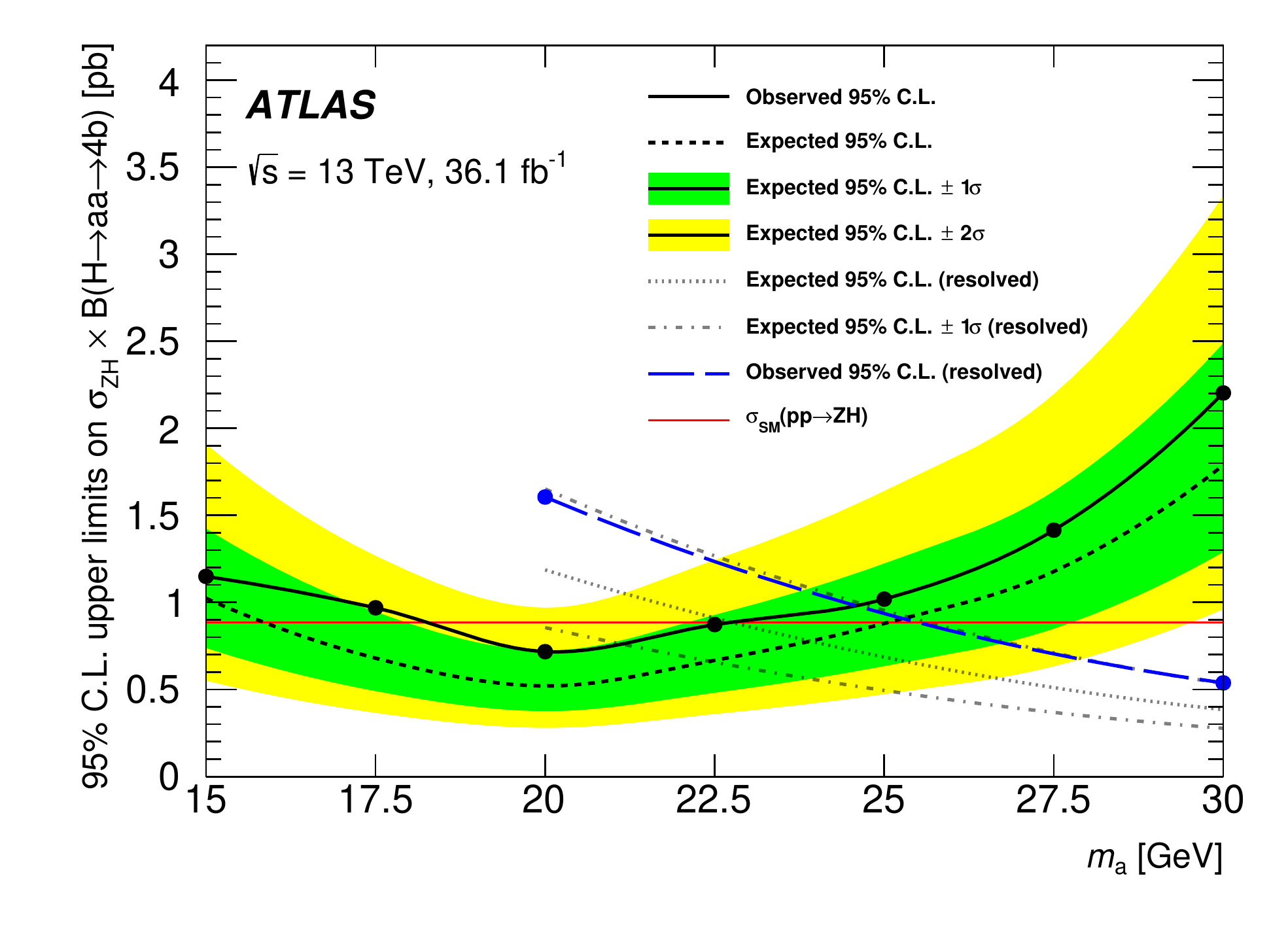}
\caption{Upper limits on $B(H\rightarrow aa \rightarrow
2b2\mu)$ obtained with 139 \ifb of data \protect\cite{H2b2m_ATL}
(left) and $\sigma_{ZH} \times B(H\rightarrow aa \rightarrow 4b)$ obtained with 36.1 \ifb of data \protect\cite{H4b_ATL} (right) as a function of the signal mass. Black and red dots show
masses for which the hypothesis testing was done. Blue lines show
comparisons with previously published ATLAS results based on  36 \ifb of data: $H\rightarrow aa
\rightarrow 2b2\mu$ search \protect\cite{H2b2mOld_ATL} (left) and $H\rightarrow aa \rightarrow 4b$
search in the final state with four resolved $b$-jets \protect\cite{H4bOld_ATL} (right).}
\label{fig:bbmm2}
\end{figure}

\subsection{$H\rightarrow aa \rightarrow 4b$}
\label{subsec:H4b}

The ATLAS search for boosted pseudoscalars in the 15 $<m_a<$ 30 GeV
mass range~\cite{H4b_ATL} developed specialised techniques to tag low-mass $a\ra bb$
objects. The analysis uses lepton triggers to target events where the Higgs boson is produced
in association with a $Z$-boson and $Z$-boson decays leptonically.
 The $a\ra bb$ reconstruction procedure starts by 
reclustering  jets with the standard radius parameter of $R=0.4$ into larger $R=0.8$ jets. Tracks matched to $R=0.8$ jets
are clustered into two or three sub-jets using the Ex$k_t$~\cite{exkt}
algorithm. With this procedure, $b$-quarks are associated to different
sub-jets with nearly 100\% efficiency. A BDT discriminant is trained
to distinguish $a\ra bb$ from the background of jets containing only
one $b$-quark. Multivariate $b$-tagging scores of the sub-jets, the
angular separation between the sub-jets and the \pt~asymmetry, (\pt$^1-$\pt$^2)/($\pt$^1+$\pt$^2)$, are used in the training. Two working
points of the tagger are calibrated in the data using gluon$\ra bb$
events and used to define a set of control and signal regions with
different signal versus background contributions and different
background compositions. The upper limits  on $B(H\rightarrow aa \rightarrow
4b)$ are shown in Figure~\ref{fig:bbmm2}, right. Below $m_a=25$ GeV, there is a
significant gain in sensitivity compared to the resolved analysis~\cite{H4bOld_ATL}
shown in blue. At around $m_a=20$ GeV, the boosted analysis is
probing the region below $B(H\rightarrow aa \rightarrow
4b)<100$\%, setting a solid ground for future improvements.


\subsection{$H\rightarrow aa \rightarrow 2\mu2\tau$}
\label{subsec:Hmmtt}
CMS searched for \Haa  in the $3.6<m_a<21$
GeV mass range in the final state with two muons ($a\ra \mu\mu$), one
hadronically decaying $\tau$ and one leptonically
(muon) decaying $\tau$, $a\ra \tau_{h} \tau_{\mu}$~\cite{H2m2t_CMS}. In this mass
range, the two $\tau$-leptons are collimated and a specialised
reconstruction technique is developed in order to reconstruct the
$a\ra \tau_{h} \tau_{\mu}$ object. Standard hadron-plus-strips (HPS)
algorithm~\cite{HPS} for reconstruction of $\tau_{h}$ candidates is modified in
two ways. Firstly, muons with \pt~$ >3$ GeV are removed from jets seeding
the HPS algorithm, resulting in a significant  increase in $a\ra
\tau_{h} \tau_{\mu}$ reconstruction efficiency at lower masses (see
Figure~\ref{fig:2m2t}). Secondly, the muon energy is excluded from the $\tau_{h}$
isolation discriminant, resulting in an efficiency increase over the
whole mass range. Standard Model backgrounds are constrained in a 2D
unbinned fit to the dimuon and four-object ($m_{\mu\mu \tau_{h}
  \tau_{\mu}}$) invariant masses. The background model fits the
data well and no significant excess above the SM predictions is
observed. The upper limits on  $B(H\rightarrow aa \ra 2\mu2\tau)$ are
show in Figure~\ref{fig:2m2t}.

\begin{figure}
\includegraphics[width=0.56\linewidth]{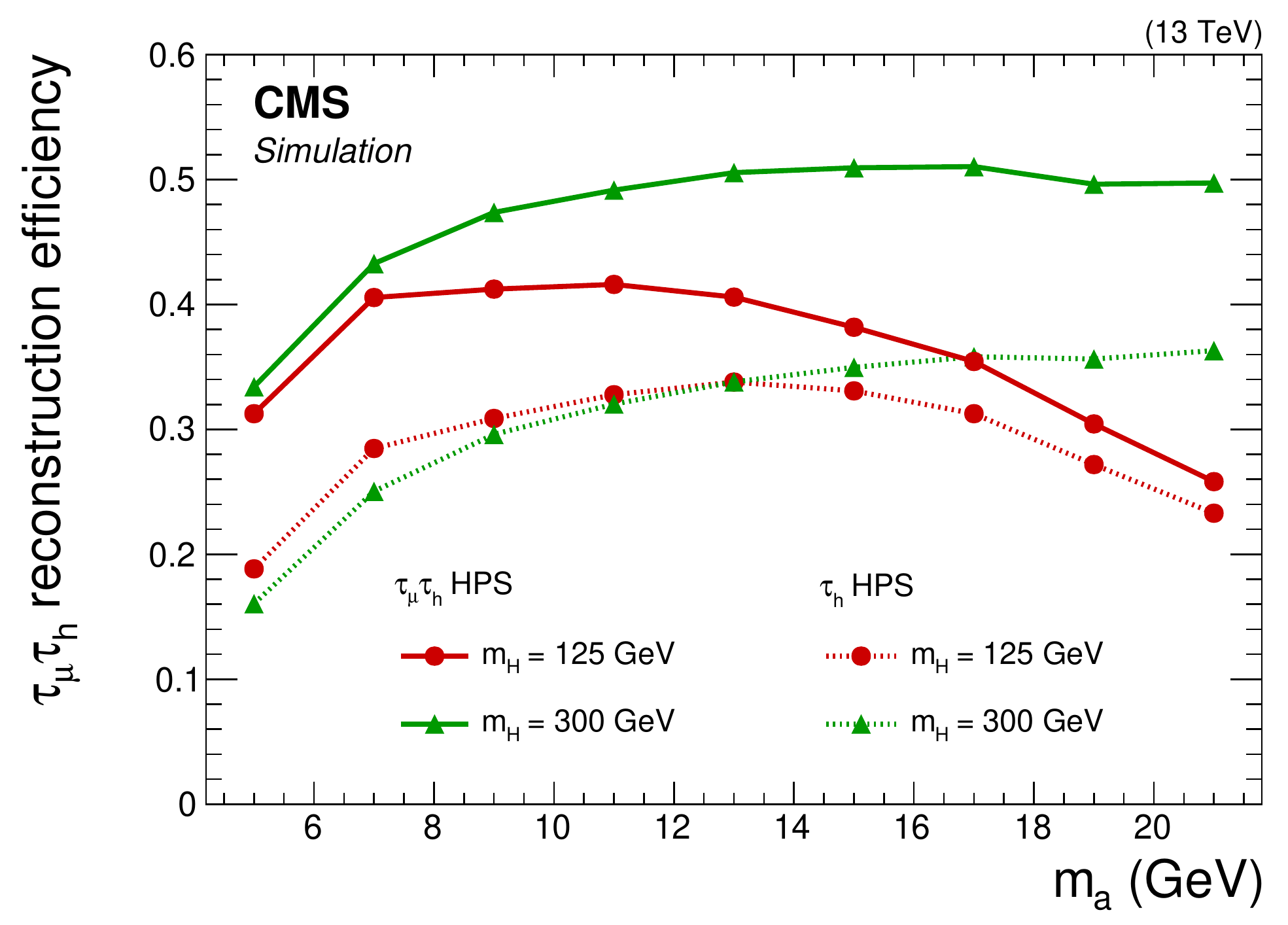}
\includegraphics[width=0.42\linewidth]{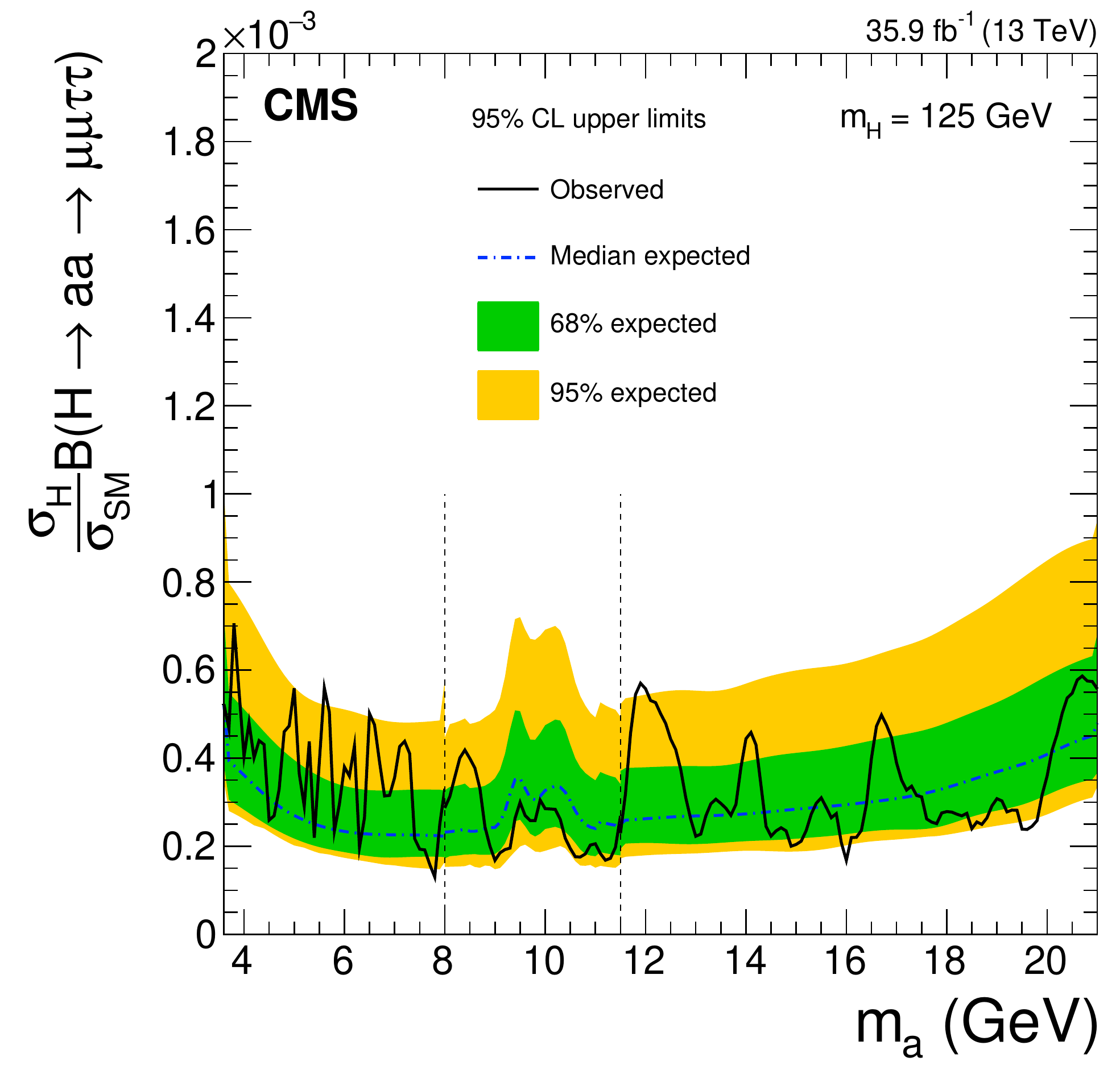}
\caption{Left: The efficiency of the standard HPS (dashed lines) and
  $\tau_{h} \tau_{\mu}$ HPS reconstruction used in $H\rightarrow aa
  \rightarrow 2\mu2\tau$ analysis \protect\cite{H2m2t_CMS} (solid lines) as a function of
  $m_a$. Right: Upper limits on $\frac{\sigma_H}{\sigma_{SM}}
  B(H\rightarrow aa \ra 2\mu2\tau )$  as a function of the signal mass.}
\label{fig:2m2t}
\end{figure}

\subsection{$H\rightarrow aa$ summary plots}
Model independent limits on $B(H\rightarrow aa \ra xx yy)$, where $x$
and $y$ represent various final-state particles, are translated into
limits on $B(H\rightarrow aa)$ under an assumption of a particular
2HDM+S scenario that specifies $B(aa \ra xx yy)$. Figure~\ref{fig:summ}
shows interpretations of ATLAS and CMS $H\rightarrow aa$ searches~\cite{Haa_ATL,Haa_CMS} in
the Type-III 2HDM+S scenario~\cite{Curtin,Uli}, with the ratio of vacuum expectation values
of the two Higgs doublets tan$\beta$=2. 
The plots demonstrate how different final states dominate the sensitivity at different $m_a$. In
the full mass range the observed limits are below the current upper limit
on $B(H\ra$ undetected) of 19\%, i.e. direct searches for exotic
Higgs decays are probing so far unconstrained phase space.

\begin{figure}
\centering
\includegraphics[width=0.54\linewidth]{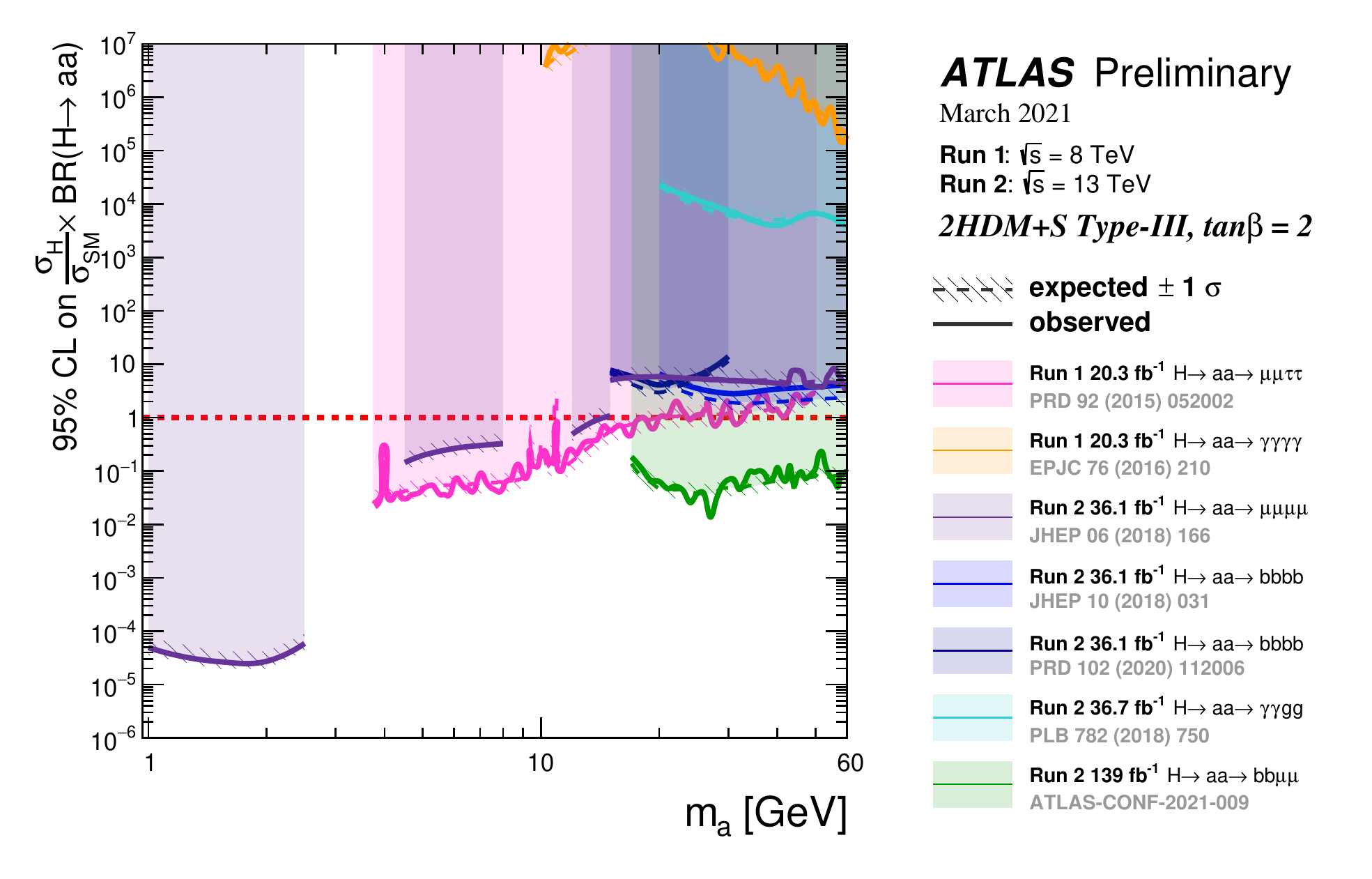}
\includegraphics[width=0.44\linewidth]{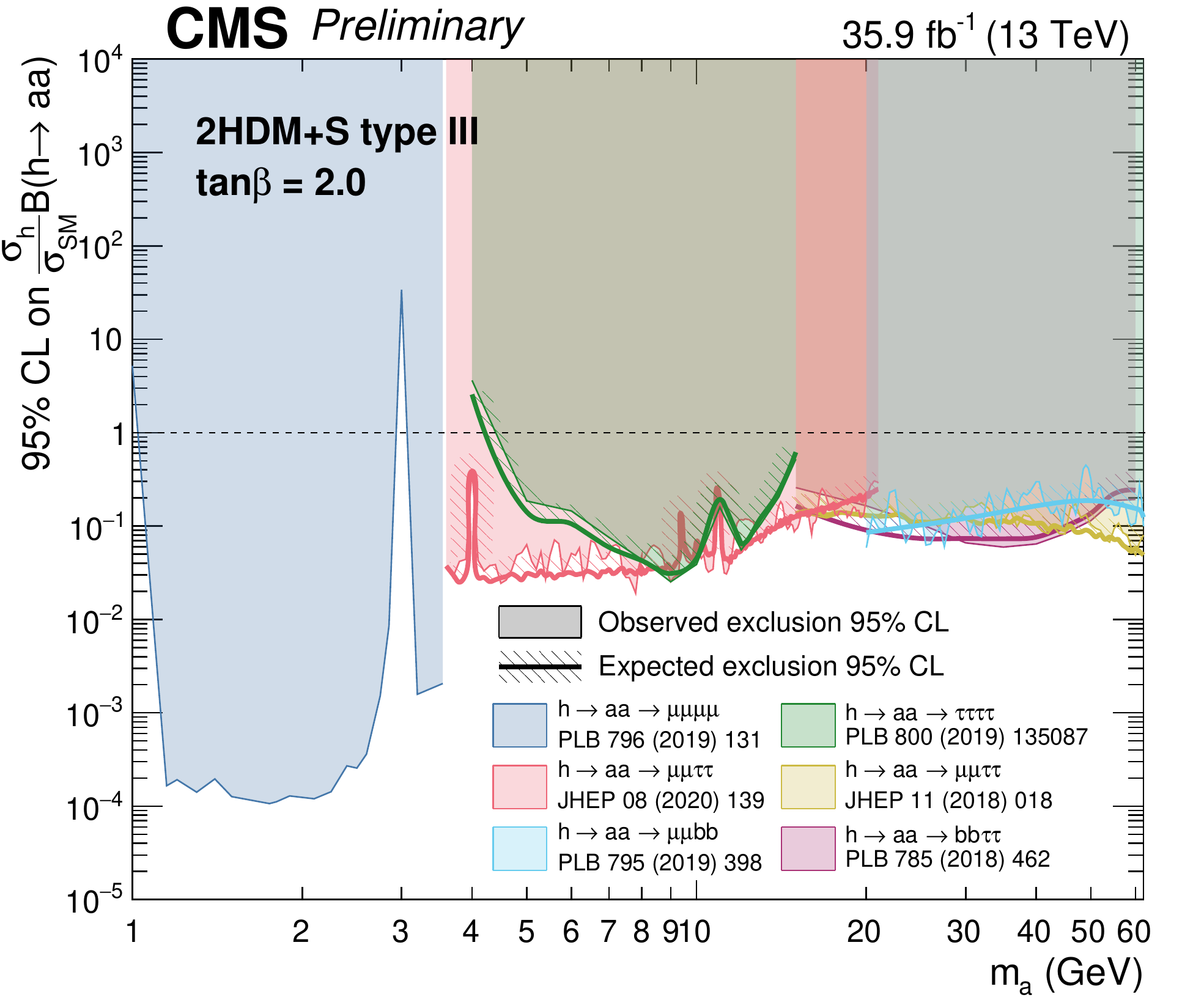}
\caption{Overlay of observed and expected upper limits on $\frac{\sigma_H}{\sigma_{SM}} B(H\rightarrow aa)$ in Type-III 2HDM+S
  scenario with tan$\beta$=2 for ATLAS \protect\cite{Haa_ATL} (top) and CMS \protect\cite{Haa_CMS} (bottom).  }
\label{fig:summ}
\end{figure}

\section{Lepton-flavor violating decays}
\subsection{$H\rightarrow  \mu\tau / e\tau$}

Lepton-flavour violating decays of the Higgs boson are searched for by
CMS in four channels: $\mu\tau_h, \mu\tau_e, e\tau_h$ and
$e\tau_{\mu}$~\cite{LFV_CMS}. The standard HPS algorithm is used to reconstruct
$\tau_h$. A BDT discriminant is trained in each channel
separately. Typical input variables to the training, among others, are lepton
transverse momenta, $\Delta\phi$ separation between the leptons and
the missing transverse momentum and the transverse mass.
Each channel is further divided into four event categories, based on
the number of jets in the final state and the di-jet invariant mass, to enhance 
different Higgs production mechanisms. 
A simultaneous fit to the BDT distributions is
performed over all channels and categories to extract
upper limits on $B(H\rightarrow  \mu\tau)$ and $B(H\rightarrow
e\tau)$. The result is shown in Figure~\ref{fig:LFV2}. The upper limits set on $B(H\rightarrow  \mu\tau)$ and $B(H\rightarrow
e\tau)$ amount to 0.15\% and 0.22\%, respectively, and improve
previous ATLAS and CMS results by about a factor two.

\begin{figure}[H]
\centering
\includegraphics[width=0.92\linewidth]{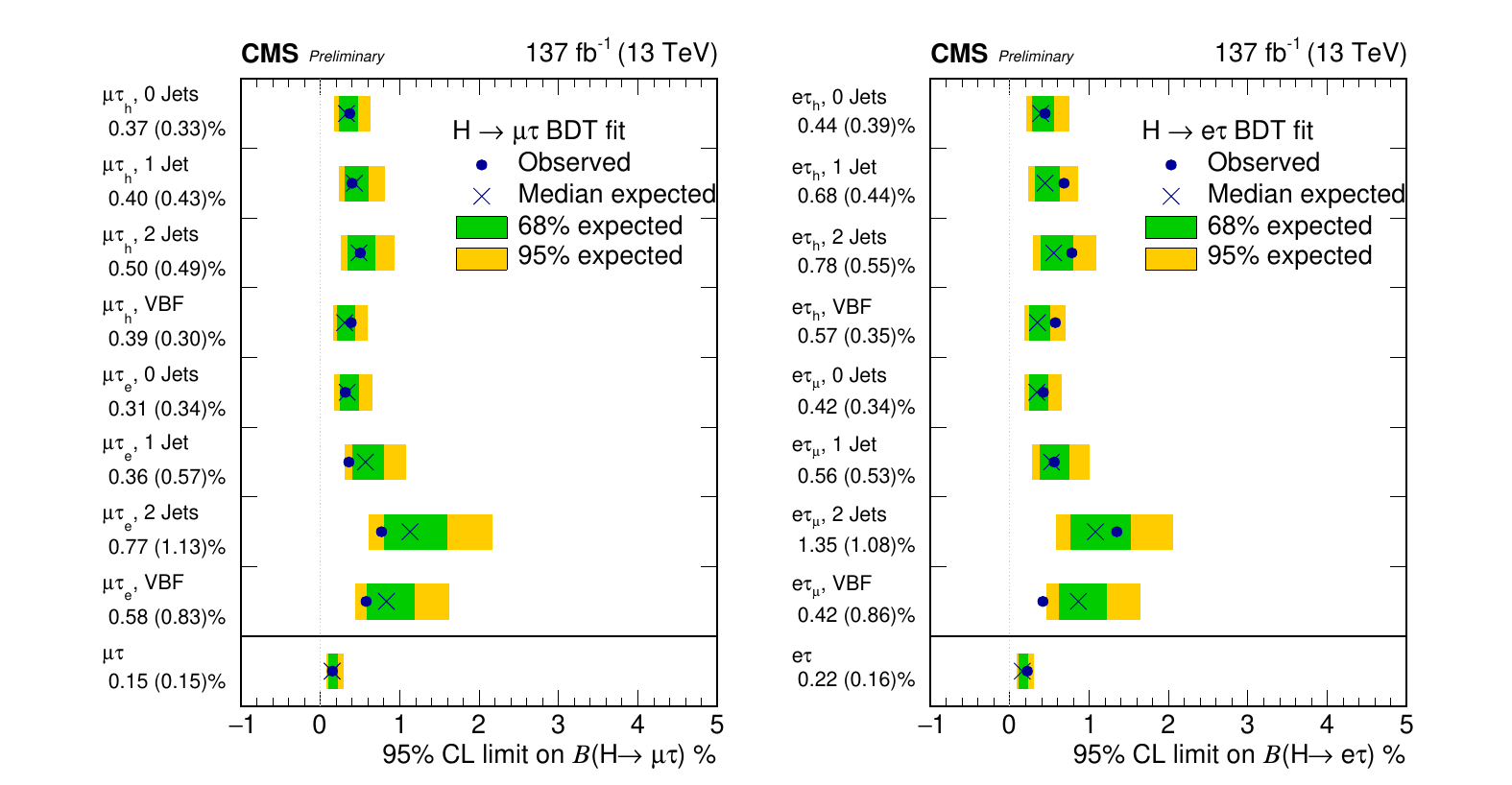}
\caption{Observed and expected upper limits on lepton-flavor
  violating Higgs branching fractions, $B(H\rightarrow
  \mu\tau)$ (left) and $B(H\rightarrow
e\tau)$ (right), shown for each event category separately and for all
the categories combined \protect\cite{LFV_CMS}.}
\label{fig:LFV2}
\end{figure}

\section{Rare decays}
\subsection{$H\rightarrow \ell\ell \gamma$}
ATLAS searched for the Higgs boson decay into two leptons (muons or
electros) and a photon, in the kinematic range where the dilepton
mass, $m_{\ell\ell}$, is below 30 GeV~\cite{Hlly_ATL}. In this mass range, the
analysis is mainly sensitive to dilepton production through
$\gamma^*$. Multiple event categories, such as $ee/\mu\mu$ resolved,
$ee$ merged, VBF enhanced, are treated separately 
to maximise the analysis sensitivity. A dedicated technique to
reconstruct merged electrons is developed to improve the efficiency in this category.
The $H\rightarrow \ell\ell \gamma$ signal is observed
with a significance of 3.2$\sigma$, providing a first evidence of
this rare Higgs decay. Figure~\ref{fig:lly} shows the 3-object invariant mass,
$m_{\ell\ell\gamma}$, spectrum, with the combined
signal-plus-background model fit to all analysis categories
simultaneously shown with the red curve.
The best-fit value of the signal-strength parameter, defined as the ratio of the
observed signal yield to the signal yield expected in SM, is measured
to be $\mu = 1.5 \pm 0.5$.

\begin{figure}
\centering
\includegraphics[width=0.49\linewidth]{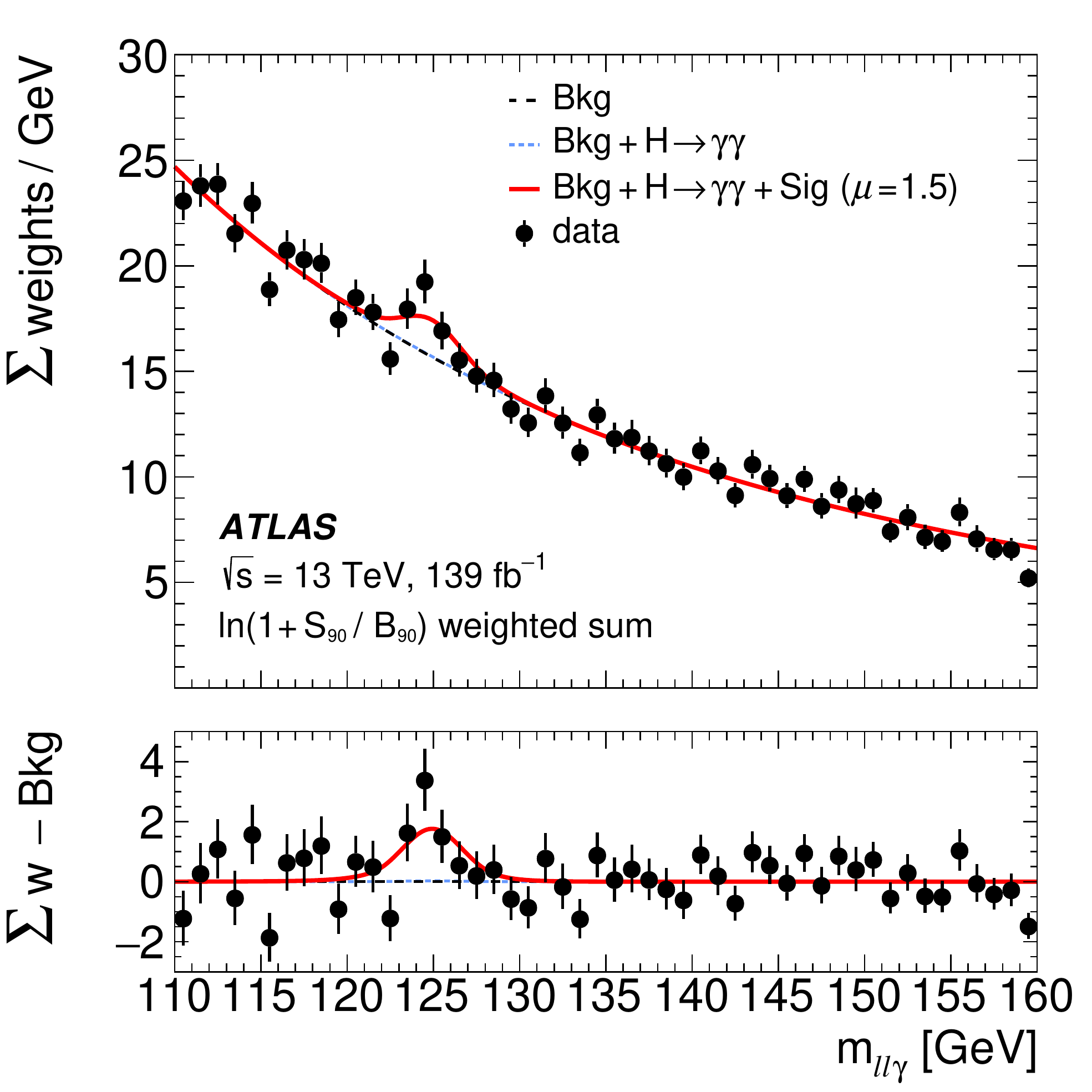}
\caption{The $m_{\ell\ell\gamma}$ distribution from all the event
  categories considered in the $H\rightarrow \ell\ell \gamma$ search \protect\cite{Hlly_ATL},
  added with a category-dependent weight. The red curve shows the
  combined signal-plus-background model. The bottom panel shows the residuals of the data with respect to the non-resonant background component of the signal-plus-background fit.}
\label{fig:lly}
\end{figure}

\section{Summary}

This presentation summarised several recent searches for rare and exotic Higgs boson decays
carried out by  the ATLAS and CMS collaborations.
The presented analyses are based on the data from 13 TeV proton-proton
collisions at the LHC. The searches for Higgs
boson decays into BSM particles and lepton-flavor violating Higgs
decays continue to probe so-far unconstrained parameter space. In the
domain of rare Higgs decays, evidence for the rare $H\rightarrow
\ell\ell \gamma$ decays was found in ATLAS, amounting to an observed
significance of $3.2\sigma$.

\end{document}
